\documentclass[preprint,showpacs,preprintnumbers,amssymb,superscriptaddress,aps,prd,nofootinbib,11pt]{revtex4-1}

\selectfont 

\usepackage{graphicx}       
\usepackage{dcolumn}        
\usepackage{bm}             
\usepackage{psfrag}
\usepackage{tensor}
\usepackage[usenames]{color}
\definecolor{navyblue}{rgb}{0.0, 0.0, 0.5}
\usepackage[linktocpage,colorlinks=true,allcolors=navyblue]{hyperref}
\usepackage{amsmath}
\usepackage{amsfonts}
\usepackage{caption}
\usepackage{subcaption}
\usepackage{physics}
\usepackage{siunitx}
\usepackage{lipsum}
\usepackage{todonotes}
\usepackage{caption}
\usepackage{subcaption}
\usepackage[skins,theorems]{tcolorbox}
\usepackage{cancel}
\tcbset{highlight math style={enhanced,
  colframe=red,colback=white,arc=0pt,boxrule=1pt}}

\begin{document}

\preprint{}

\title{Turnaround Radius for charged particles in the Reissner-Nordstr\"{o}m deSitter spacetime}

\author{Ethan J. German}
\email{ejgerman1@sheffield.ac.uk}
\affiliation{Consortium for Fundamental Physics, School of Mathematics and Statistics,
University of Sheffield, Hicks Building, Hounsfield Road, Sheffield S3 7RH, United Kingdom}
\affiliation{
Department of Mathematics, University of Malta, Msida, Malta
}
\author{Joseph Sultana}
\affiliation{
Department of Mathematics, University of Malta, Msida, Malta
}

\date{\today}

\begin{abstract}
We investigate the turnaround radius of the Reissner-Nordstr\"{o}m deSitter Spacetime and how the turnaround radius changes if a test particle carries charge. We also consider the Mart\'{i}nez-Troncoso-Zanelli (MTZ) solution of conformally coupled gravity and investigate how the turnaround radius changes for a scalar test charge. In both scalar and electric interaction cases we find that the Turnaround Radius depends on the particle's energy.
\end{abstract}

\pacs{}
\maketitle

\vspace{0.3cm}

\section{Introduction}
The concept of \textit{turnaround radius} is not new. Over the years, in the literature this has been referred to by different names, such as ``critical radius'', ``zero gravity radius'', ``maximum size of large scale structures'' and others \cite{stuchlick83,stuchlick00,mizony05,roupas14,nolan14}. The idea behind the turnaround radius is very simple. In an accelerating universe, such as the Friedmann-Lema\^{i}tre-Robertson-Walker (FLRW) model, containing a spherical inhomogeneity there is a maximum physical (aerial) distance from the centre of the inhomogeneity where a spherical shell of dust particles moving along radial timelike geodesics experience zero radial acceleration. This distance is called the turnaround radius, $r_{TR}$. In other words at this radius the gravitational attraction by the spherical inhomogeneity is exactly counterbalanced by the gravitational repulsion of the dark energy cosmological background, so that a spherical shell of dust particles just outside $r_{TR}$ having initial zero radial velocity follows the Hubble flow and expands forever, while a similar shell inside $r_{TR}$ will collapse towards the inhomogeneity. One has to point out that this critical radius does not represent an absolute boundary like for example an event horizon, in the sense that it acts as a one way membrane only for geodesic motion, i.e. a dust particle outside this radius can still cross inside if it has adequate acceleration. To obtain the turnaround radius one can either consider the radial timelike geodesics in the static geometry close to the inhomogeneity or the comoving test fluid in the expanding cosmological background \cite{riess19,aghanim18}.

In general relativity (GR) for a spherical structure of mass $M$ embedded in asymptotically de Sitter cosmological spacetime such as $\mathrm{\Lambda}$CDM, the turnaround radius is independent of the cosmic epoch and is given by $r_{TR} = (3GM/\Lambda c^2)^{1/3}$, where $\Lambda$ is the cosmological constant. This provides an upper bound on the size of the maximum structures in the observed Universe in the framework of the $\mathrm{\Lambda}$CDM model. Equivalently one can define the turnaround density $\rho_{TR} = 2\rho_{\Lambda} = 2(\Lambda c^2/8\pi G)$ as the lower bound for such structures predicted by the spherical collapse model \cite{pavlidou14}. One can therefore use $r_{TR}$ as an observable to constrain the parameters of any cosmological model by comparing the theoretical prediction of the model with actual data. For example for the $\mathrm{\Lambda}$CDM, the prediction is quite close \cite{pavlidou14,busha03,pavlidou14b} with the difference between  $r_{TR}$ and the actual size of galaxies and even super clusters as large as $M\geq10^{15}M_{\odot}$ being only about $10\%$ (see Figure 1 in \cite{pavlidou14}); although in these studies one must say that the error bars are quite large. This means that from a turnaround radius perspective the $\mathrm{\Lambda}$CDM is consistent with observations. In the last years this approach based on the turnaround radius has been used to obtain constraints for various dark energy models and modified theories of gravity, such as Brans-Dicke theory, scalar tensor gravity, $f(R)$ theory and other higher order theories \cite{faraoni16,capozziello19,lopes18,bhattacharya17,bhattacharya17b, nojiri18, hansen20}. For example in the case of Brans-Dicke theory with a cosmological constant $\Lambda$ it was found \cite{bhattacharya15,bhattacharya17} that the turnaround radius $r_{TR}$ is greater than that of $\mathrm{\Lambda}$CDM in GR. This is attributed to the fact that the scalar field in Brans-Dicke theory enhances the gravitational attractive force, which in turn produces larger structures. The effect of non-sphericity of the large scale cosmic structure on the turnaround radius was also investigated in Ref. \cite{bhattacharya21}. The turnaround radius for cosmic structures in GR can be defined more rigourously and in a gauge invariant manner (for first order perturbations of the exact FLRW model) \cite{faraoni15} in terms of the Hawking-Hayward quasi-local energy \cite{hawking68,hayward94,hayward96}. In this approach the quasi-local mass of the cosmic structure is divided into a local and a cosmological component in order to facilitate the interplay between the local attraction and cosmic expansion. The value for the turnaround radius obtained with this approach is quite similar to that obtained using the standard approach of geodesic motion as described above. It was also shown \cite{lapierre17} that alternative definitions of quasi-local mass in GR \cite{szabados09}, such as that of Brown and York \cite{brown93}, leads to the same expression for the turnaround radius in the case of first order cosmological perturbations. The concept of quasi-local energy has been generalized to scalar-tensor theories \cite{cai09,cai08,wu08,cognola11,faraoni16b,hammad16}, but not to more general modified gravity theories and so in this case this approach cannot be used \cite{nojiri18}.

In all the earlier studies the turnaround radius was obtained by considering free neutral test particles. In this study we obtain the turnaround radius for the Reissner-Nordstr\"{o}m-de Sitter (RNdS) spacetime and for the hairy black hole (with a conformally coupled scalar field) in a cosmological background, obtained by Mart\'{i}nez, Troncoso and Zanelli (MTZ)\cite{MTZ}. This is done by analyzing the motion of a test electric charge and a test scalar charge in these spacetimes respectively. Unlike a neutral free particle, the non-geodesic motion of these test charges depends on the electric charge $Q$ and scalar field $\phi$ in these spacetimes and so this has an effect on the position of the turnaround radius. The structure of this paper is as follows. In the next section we review the calculation of the turnaround radius for spherically symmetric spacetimes. In the following section we obtain the equation of motion for a test electric charge in the RNdS spacetime and for a test scalar charge in the hairy black hole spacetime obtained by Martinez et al. The main results of these analyses are presented and discussed in Section IV. The paper ends with a Conclusion in Section V. In this article we use geometric units in which $G=c=1$ and we take the value of the cosmological constant $\Lambda=1.1\times 10^{-52} \unit{m^{-2}}$. Dots over variables refer to derivatives with respect to proper time $\tau$, whereas primes refer to derivatives with respect to an arbitrary parameter of the worldline $\lambda$. 

\section{Turnaround Radius of a spherically symmetric spacetime}
The turnaround radius of a spacetime is defined as the outermost radius at which:
\begin{equation}
    \dv[2]{R}{\tau} = 0,
\end{equation}
where $\tau$ is proper time and $R$ is the \emph{areal radius} of the spacetime defined $R=\sqrt{\frac{A}{4\pi}}$ where $A$ is the area of the 2-sphere of symmetry in the spacetime. This definition is entirely equivalent to the following condition:
\begin{equation}
\dv{V}{r} = 0 ,
\end{equation}
where $V$ is the effective potential for timelike radial trajectories in this spacetime. We consider spherically symmetric spacetimes of the form:
\begin{equation}
    \dd s^2 = - A(r) \dd t^2 + \frac{1}{A(r)}\dd r^2 + r^2 \dd \Omega^2,
\end{equation}
where $\dd \Omega ^2 = \dd \theta^2 + \sin^2(\theta) \dd \phi^2$. Spacetimes of this form have areal radius equal to the radial coordinate. This metric has two symmetries of interest, these symmetries correspond to the constants of motion $E$ and $L$, i.e., the energy and angular momentum  per unit mass respectively. Such spacetimes will have effective potential given by:
\begin{equation}
    V(r) = \frac{1}{2}\left(\frac{L^2}{r^2} +\epsilon\right)A(r),
\end{equation}
where $\epsilon$ is +/- $1$ for timelike/spacelike geodesics and 0 for null geodesics. For timelike radial trajectories we have that $L=0$ and $\epsilon=1$, which reduces the effective potential to $V(r) = \frac{1}{2} A(r)$. Hence the condition for defining the turnaround radius for spherically symmetric spacetimes is $A'(r)=0$. Applying this to the Schwarzchild-de Sitter spacetime which has $A(r) = 1-\frac{2M}{r} -\frac{\Lambda}{3}r^2$ we get the standard turnaround radius found in the literature\cite{Faraoni_2015,pavlidou14}:
\begin{equation}
    r_{TR} = \left(\frac{3M}{\Lambda}\right)^{\frac{1}{3}}.
\end{equation}
Applying this to the RNdS spacetime which has $A(r) = 1-\frac{2M}{r} - \frac{\Lambda}{3} r^2 + \frac{Q^2}{r^2}$, where $Q$ is the charge of the central black hole, we find that the corresponding turnaround radius is one of the roots of the following polynomial:
\begin{equation}
    P(r) = \Lambda r^4 - 3 Mr + 3Q^2. \label{eqn:TRRNdSPoly}
\end{equation}
\section{Turnaround Radius for particles interacting with the spacetime}
\subsection{Electrically charged particle interaction}
Consider a particle with charge-mass ratio $\epsilon = q/\mu$ in the RNdS spacetime. The motion of such a particle will be governed by the following action principal \cite{chandrasekhar1998mathematical, chargedparticlemotion}:
\begin{equation}
    S = \int \dd \tau \ \left\{-\frac{1}{2} g_{\mu\nu}\dot{x}^\mu\dot{x}^\nu + \epsilon A_\mu \dot{x}^\mu\right\},\label{eqn:ChargedParticleAction}
\end{equation}
where $A^\mu$ is the electromagnetic vector-potential with only one non-zero component $A_0 =- Q/r$. The associated equations of motion are found by varying this action with respect to the coordinates, giving \cite{chargedparticlemotion}:
\begin{equation}
     \frac{D^2 x^\mu}{\dd^2 \tau} = \epsilon F^\mu_{\ \ \nu \ } \dot{x}^\nu,
\end{equation}
where:
\begin{equation}
    \frac{D^2 x^\mu}{\dd^2 \tau} = \ddot{x}^\mu + \Gamma^\mu_{\ \alpha \beta} \dot{x}^\alpha \dot{x}^\beta,
\end{equation}
and $F_{\mu\nu} = A_{\mu,\nu}-A_{\mu,\nu}$ is the Maxwell tensor. Such equations of motion require that the the tangent vector is a unit timelike vector, hence $g_{\mu\nu} \dot{x}^\mu\dot{x}^\nu = -1$. The Lagrangian in Eq.~\eqref{eqn:ChargedParticleAction} is cyclic in $t$ and $\phi$, and hence we have the following conserved quantities:
\begin{align}
    A(r) \dot{t}+\frac{\epsilon Q}{r} =  E, \qquad r^2 \dot{\phi} = L,
\end{align}
with $E$ and $L$ defined as before. Again, we consider radial trajectories so $L=0$, allowing us to derive the following effective potential from $g_{\mu\nu}\dot{x}^\mu\dot{x}^\nu =-1$:
\begin{equation}
    V(r) = \frac{1}{2}\left\{A(r) + 2 \frac{\epsilon E Q}{r} - \frac{\epsilon^2 Q^2}{r^2}\right\}. \label{eqn:chargedParticleEffective Potential}
\end{equation}
Applying the same procedure as above, we find that the turnaround radius for charged particles is the maximal root of the following polynomial:
\begin{equation}
    P^\epsilon(r) = \Lambda  r^4 +3 rE Q \epsilon -3 Mr -3 Q^2 \left(\epsilon ^2-1\right).  \label{eqn:TRChargedParticle}
\end{equation}
Notice in the charge-less limit this reproduces the result in Eqn.~\eqref{eqn:TRRNdSPoly}.
\subsection{Scalar field interaction}
\subsubsection{Scalar Tensor Theory and the MTZ solution}
We now consider a black hole in a scalar tensor theory of gravity. The action principal for the conformally coupled theory of gravity with cosmological constant ($\Lambda$CCG) is:
\begin{equation}
    S^{\Lambda \mathrm{CCG}}=\int \mathrm{d}^4 x \sqrt{-g}\left[\frac{R-2 \Lambda}{16 \pi }-\frac{1}{2} g^{\mu \nu} \partial_\mu \varphi \partial_\nu \varphi-\frac{1}{12} R \varphi^2-\alpha \varphi^4\right],
\end{equation}
where $\varphi$ is the scalar field that is non-minimally coupled to gravity through a higgs-like potential, and $\alpha$ is a dimensionless constant. Varying the action with respect to the metric tensor and the scalar field gives field equations: 
\begin{align}
    G_{\mu\nu} + \Lambda g_{\mu\nu} = 8\pi T_{\mu\nu} \text{, }\\
    \square \varphi-\frac{1}{6} R \varphi-4 \alpha \varphi^3=0 \text {, }
\end{align}
where the stress tensor is given by:
\begin{equation}
    T_{\mu v}=\partial_\mu \varphi \partial_\nu \varphi-\frac{1}{2} g_{\mu v} g^{\alpha \beta} \partial_\alpha \varphi \partial_\beta \varphi+\frac{1}{6}\left[g_{\mu \nu} \square-\nabla_\mu \nabla_\nu+G_{\mu v}\right] \varphi^2-\alpha g_{\mu \nu} \varphi^4 .
\end{equation}
These field equations are invariant under conformal transformations $g_{\mu \nu} \rightarrow \Omega^2(x) g_{\mu \nu},\ \varphi \rightarrow \Omega^{-1}(x) \varphi$.

A black hole solution to this theory of gravity was reported by Mart\'{i}nez et.al. \cite{MTZ} given by the metric induced by the line element:
\begin{equation}
    \mathrm{d} s^2=-\left[-\frac{\Lambda}{3} r^2+\left(1-\frac{M}{r}\right)^2\right] \mathrm{d} t^2+\left[-\frac{\Lambda}{3} r^2+\left(1-\frac{M}{r}\right)^2\right]^{-1} \mathrm{~d} r^2+r^2 \mathrm{~d} \Omega^2,
\end{equation}
together with scalar field:
\begin{equation}
    \varphi(r)=\sqrt{\frac{3}{4 \pi}} \frac{M}{r-M}.
\end{equation}
This is only a solution for $\alpha = -\frac{2}{9}\pi \Lambda$. This metric is identical to the extreme RNdS solution, i.e. the RNdS metric with $Q=M$ \cite{MTZ}. Thus the turnaround radius of this spacetime for particles that do not interact with the scalar field follows directly from Eq.~\eqref{eqn:TRRNdSPoly} by setting $Q=M$.
\subsubsection{Motion of scalar test charges}

In this section we summarise the setup used by Bekenstein ~\cite{BekensteinJacobD1975Bhws} to consider the motion scalar test charges moving in a spacetime with a scalar field coupled to the gravitational interaction. We consider scalar test charges having rest mass $\mu$ and coupling strength $f$ with the main scalar field $\varphi$ of the MTZ spacetime. The motion of these test charges are described by the following Lagrangian: 
\begin{equation}
    \mathcal{L} = -(\mu + f \varphi) \sqrt{-g_{\mu\nu} x'^\mu x'^\nu}.
\end{equation}
The equations of motion are obtained by varying the corresponding action with respect to  the coordinates and are given by:
\begin{equation}
    \frac{D^2x^\nu}{\dd \lambda} = -\mu^{-2}f(\mu+f\varphi) \nabla^\nu\varphi,
\end{equation}
where the parameter $\lambda$ along the trajectories is chosen such that
\begin{equation}
    -g_{\mu\nu}x'^\mu x'^\nu =  \mu^{-2}(\mu+f\varphi)^2.  \label{eqn:parameterisation}
\end{equation}
Note that if $\lambda$ is the proper time $\tau$ then $-g_{\mu\nu}x'^\mu x'^\nu =1$. Thus the relationship between $\lambda$ and $\tau$  is given by:
\begin{equation}
    \left(\dv{\tau}{\lambda}\right)^2 = \frac{(\mu + f \varphi)^2}{\mu^2}. \label{eq:dtaudalabdasquared}
\end{equation}
Now, $\mathcal{L}\neq \mathcal{L}(t)$ therefore $-E=\pdv{\mathcal{L}}{t'}$ is a conserved quantity. Evaluating this for the MTZ spacetime gives:
\begin{equation}
t' = -\frac{E}{\mu A(r)}.
\end{equation}
$\mathcal{L}$ is also independent of $\phi$ thus $L=\pdv{\mathcal{L}}{\phi'}$ is also a conserved quantity, however here we are only concerned with radial trajectories so $L=0$. Now by Eq.~\eqref{eqn:parameterisation} one can derive the following expression for $r'^2$:
\begin{equation}
    r'^2 = - \mu^{-2}A(r) (\mu+f\varphi)^2 + E^2/\mu^2, \label{eqn:rprimesqr}
\end{equation}
giving the effective potential:
\begin{equation} 
    V(r) = \frac{1}{2}\mu^{-2} A(r) (\mu + f\varphi)^2.
\end{equation}
\subsubsection{The turnaround radius for scalar test charges}
To find the position of the turnaround radius we cannot simply set the derivative of the effective potential $V(r)$ to zero as this would correspond to $r'' = 0$. Since the position of the turnaround radius corresponds to $\ddot{r} =0$ (where dot denotes differentiation with respect to proper time) we use the chain rule to write:
\begin{equation}
    \ddot{r} = \left(\dv{\lambda}{\tau} \right)^2 r'' + r' \dv[2]{\lambda}{\tau}.
\end{equation}
From Eq.~\eqref{eq:dtaudalabdasquared} one can calculate the second derivative term:
\begin{equation}
    \dv[2]{\lambda}{\tau} = -f \dv{\varphi}{r} r' \dv{\lambda}{\tau} \frac{\mu}{(\mu +f\varphi)^2}.
\end{equation}

Hence $\ddot{r} = 0$ implies that 
\begin{equation}
    r'' = (r')^2 f \dv{\varphi}{r}.\frac{1}{\mu + f \varphi}
\end{equation}
Now, we use that $r'' = -\dv{V}{r}$ and Eq.~\eqref{eqn:rprimesqr}, to find that the turnaround radius is the maximal root of the polynomial:
\begin{align}
    \nonumber P^f(r) &=3 \sqrt{\frac{3}{\pi }} f^3 M^3 P(r)-18 f^2 \mu  M^2 (M-r) P(r)-8 \pi  \mu ^3 (M-r)^3 P(r)\\
    &+12 \sqrt{3 \pi } f \mu ^2 M (M-r) \left(3 M^3-6 M^2 r+M r^2 \left(\Lambda  r^2+3\right)-r^3 \left(E^2+\Lambda  r^2\right)\right), \label{eqn:TRScalarCharge}
\end{align}
where $P(r)$ is the polynomial in Eq.~\eqref{eqn:TRRNdSPoly}. In the limit when $f\rightarrow0$ this polynomial clearly reduces to a condition equivalent to finding the roots of $P(r)$. One thing to note is that this polynomial, and hence its roots, depends on the parameters $\mu, f,$ and notably $E$.
\section{Analysis of Results}

In this section we numerically analyse the results. We first develop approximations to the positions of the roots of the polynomials. We then illustrate how the turnaround can depend on the field interactions. Finally we compute radial trajectories starting close to the turnaround radius numerically to show that the the position we find is indeed the turnaround radius. 

\subsection{Asymptotic approximation of roots.}
To find the turnaround radius given the polynomials defined in equations Eq.~\eqref{eqn:TRRNdSPoly}, Eq.~\eqref{eqn:TRChargedParticle} and Eq.~\eqref{eqn:TRScalarCharge} one can plug in values for the parameters and use standard root finding algorithms available in software, such as using Mathematica's \verb|NSolve[]|. However, this only returns the values of the roots for the given parameters and doesn't tell us anything more useful about the nature of the roots. Luckily, these polynomials are all functions of $\Lambda$ which is a very small (on the order of $10^{-52}$m$^{-2}$). Because of this we can define a perturbation expansion in $\Lambda$ to obtain closed form approximations to the roots. Details on these root's calculation can be found in Appendix \ref{appendix:PertExp}. Here we report the results and analyse the expressions.

\subsubsection{Uncharged interaction case.}
In the uncharged case the turnaround radius is given by the roots of the polynomial in Eq.~\eqref{eqn:TRRNdSPoly}. Here we have dependence on black hole mass $M$, black hole charge $Q$ and the cosmological constant $\Lambda$. This polynomial is of degree 4, and thus should have 4 roots. Writing the roots as a linear perturbation expansion in $\Lambda$ gives the following expressions for the roots of the polynomial:
\begin{align}
    r^\text{TR}_1 &=  \frac{Q^2}{M} + \mathcal{O}(\Lambda^{1/3})\\
    r^\text{TR}_2 &= -\frac{Q^2}{3 M} -\frac{1}{2}(1+i\sqrt{3}) \sqrt[3]{\frac{3M}{\Lambda }} + \mathcal{O}(\Lambda^{1/3})\\
    r^\text{TR}_3 &= -\frac{Q^2}{3 M} - \frac{1}{2}(1-i\sqrt{3})\sqrt[3]{\frac{3M}{\Lambda }} + \mathcal{O}(\Lambda^{1/3})\\
    r^\text{TR}_4 &= -\frac{Q^2}{3 M}+\sqrt[3]{\frac{3M}{\Lambda }}+ \mathcal{O}(\Lambda^{1/3})
\end{align}
where $i = \sqrt{-1}$. $r^\text{TR}_1$ will always be inside or at the event horizon $r=M+\sqrt{M^2 -Q^2}$. $r^\text{TR}_2$ and $r^\text{TR}_3$ are very clearly a complex conjugate pair. This leaves $r^\text{TR}_4$ as the only remaining physical root. One can clearly see that in the $Q\rightarrow0$ limit of $r^\text{TR}_4$ the Schwarzschild  de Sitter turnaround radius is obtained. Thus in the uncharged case the only possible value for the Turnaround Radius is:
\begin{equation}
    r^\text{TR}= -\frac{Q^2}{3 M}+\sqrt[3]{\frac{3M}{\Lambda }}+ \mathcal{O}(\Lambda^{1/3}). \label{eqn:TRRNdS}
\end{equation}
\subsubsection{Electrically charged interaction case.}
In the electrically charged case the turnaround radius is given by the roots of the polynomial in Eq.~\eqref{eqn:TRChargedParticle}. Here we have dependence on black hole mass $M$, black hole charge $Q$, the particle's charge $\epsilon$, and the constant of motion $E$. As before, this is a polynomial of degree 4 and thus should have 4 roots. Expanding in a linear perturbation in $\Lambda$ gives the following expressions for the 4 roots:
\begin{align}
    r^\text{TR}_1 &= \frac{Q^2 (\epsilon^2 -1)}{E Q \epsilon -M} + \mathcal{O}(\Lambda^{1/3})\\
    r^\text{TR}_2 &= \frac{Q^2 (\epsilon^2 -1)}{3 (M-E Q \epsilon )}-\frac{1}{2} (1+i\sqrt{3})\sqrt[3]{\frac{3( M-E Q \epsilon) }{\Lambda }} + \mathcal{O}(\Lambda^{1/3})\\
    r^\text{TR}_3 &= \frac{Q^2 (\epsilon^2 -1)}{3 (M-E Q \epsilon )}-\frac{1}{2} (1-i\sqrt{3})\sqrt[3]{\frac{3( M-E Q \epsilon) }{\Lambda }} + \mathcal{O}(\Lambda^{1/3})\\
    r^\text{TR}_4 &= \frac{Q^2 (\epsilon^2 -1)}{3 (M-E Q \epsilon )}+\sqrt[3]{\frac{3( M-E Q \epsilon) }{\Lambda }} + \mathcal{O}(\Lambda^{1/3})
\end{align}
where $i = \sqrt{-1}$. $r^\text{TR}_1$ is a minimum stationary point of the effective potential. It is located outside the horizon of the RNdS black hole for certain parameters corresponding to a near extremal RNdS solution. Particles starting near this point in cases that it is outside the horizon oscillate about it. This is a known phenomena in that occurs in the RN spacetime, see refs 18-24 of \cite{chargedparticlemotion}. $r^\text{TR}_2$ and $r^\text{TR}_3$ are a complex conjugate pair and thus nonphysical roots. $r^\text{TR}_4$ is an unstable stationary point in the potential. This means that trajectories starting near the point will produce the characteristic behaviour of the Turnaround radius, and thus acts as a generalisation to Eq.\eqref{eqn:TRRNdS} in the case when the particle is electrically charged. Note that in the case when $\epsilon=\frac{M}{EQ}$ all 4 roots vanish and there is no turnaround radius. This is only possible when the sign of $\epsilon$ and $Q$ are the same. As $\epsilon\rightarrow0$ these four roots, and in particular $r_4$, reduce down to the four roots given in the uncharged case above.
\subsubsection{Scalar charged interaction case. }
In the scalar charged case the turnaround radius is given by the roots of the degree 7 polynomial in Eq.~\eqref{eqn:TRScalarCharge}.  Here we have dependence on black hole mass $M$, particle mass $\mu$, the particle's scalar field interaction strength $f$, and the energy constant of motion $E$. Expanding in a linear perturbation in $\Lambda$ gives closed form expressions for the roots:
\begin{align}
    r^\text{TR}_1 &= M + \mathcal{O}(\Lambda^{1/3})\\
    r^\text{TR}_2 &= \frac{M \left(-9 E^2 f^2+4 \sqrt{3 \pi } E^2 f \mu +4 \pi  \mu ^2\right)}{6 \mu  \left(\sqrt{3 \pi } E^2 f-2 \pi  \mu \right)}-\frac{1}{2}(1+i\sqrt{3}) \sqrt[3]{\frac{3M\mu - \frac{3}{2}\sqrt{\frac{3}{\pi}}E^2f}{\Lambda \mu}}+ \mathcal{O}(\Lambda^{1/3})\\
    r^\text{TR}_3&=\frac{M \left(-9 E^2 f^2+4 \sqrt{3 \pi } E^2 f \mu +4 \pi  \mu ^2\right)}{6 \mu  \left(\sqrt{3 \pi } E^2 f-2 \pi  \mu \right)}-\frac{1}{2}(1-i\sqrt{3})  \sqrt[3]{\frac{3M\mu - \frac{3}{2}\sqrt{\frac{3}{\pi}}E^2f}{\Lambda \mu}} \mathcal{O}(\Lambda^{1/3})\\
    r^\text{TR}_4&=\frac{M \left(-9 E^2 f^2+4 \sqrt{3 \pi } E^2 f \mu +4 \pi  \mu ^2\right)}{6 \mu  \left(\sqrt{3 \pi } E^2 f-2 \pi  \mu \right)}+ \sqrt[3]{\frac{3M\mu - \frac{3}{2}\sqrt{\frac{3}{\pi}}E^2f}{\Lambda \mu}} + \mathcal{O}(\Lambda^{1/3}) \label{eqn:TRScalarChargeRoot4}
\end{align}
Here there are only 4 roots, when a degree 7 polynomial should have 7 roots in the complex plane. The other 3 roots ($r^\text{TR}_5,r^\text{TR}_6, r^\text{TR}_7$) are listed in appendix \ref{appendix:abc}. When $f=\frac{2 M \mu }{E^2}\sqrt{\frac{\pi}{3}}$ the cubed root term in $r_2,r_3,r_4$ vanishes. With this chosen value for $f$ the first term in each case reduces to $-\frac{1}{3} M(1+3M)$ which is clearly a negative value and hence outside the domain of $r$.  In the limit as $f\rightarrow \frac{2 \mu}{E^2}\sqrt{\frac{\pi}{3}}$ the denominator of the first term in $r_4$ tends to zero. In this limit the root will tend to negative infinity except in the case when $E=1$ in which the limit is finite by l'Hospital's rule. In this case we get that:
\begin{equation}
    \lim\limits_{f\rightarrow 2\mu \sqrt{\frac{\pi}{3}}} r_{4} = \sqrt[3]{\frac{3(M-1)}{\Lambda }}-\frac{4 M}{3}
\end{equation}
\begin{figure}
    \centering
    \includegraphics{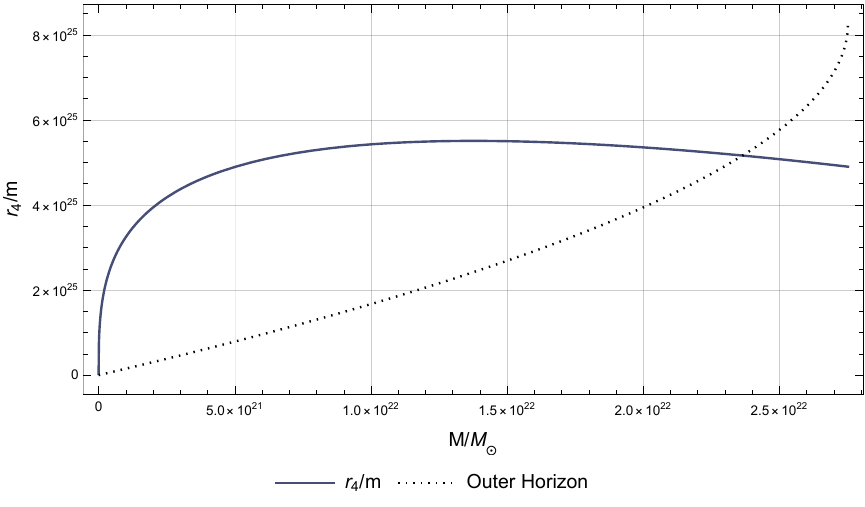}
    \caption{Mass dependence of the turnaround radius for a charge that interacts with the scalar field in the MTZ spacetime with $f= 2\mu \sqrt{\frac{\pi}{3}}$, for trajectories with $E=1$. Mass is plotted on the range $M\in [0,l/4]$ where $l= \sqrt{3/\Lambda}$, outside of this range the MTZ solution becomes a naked singularity.}
    \label{fig:massdep}
\end{figure}
In this case this limit is non-physical for $M\leq1$, however is outside of the outer horizon on a range of values $M>1$ as see in Fig.~\ref{fig:massdep}. There is also a region in the domain of $M$ where for this value of $f$ the turnaround radius is inside the outer horizon. In this case particles outside the horizon with this interaction strength will be overcome by the scalar field interaction and will shoot outward away from the black hole. 

\subsection{Dependence on field interaction}
\begin{figure}
    \centering
    \includegraphics{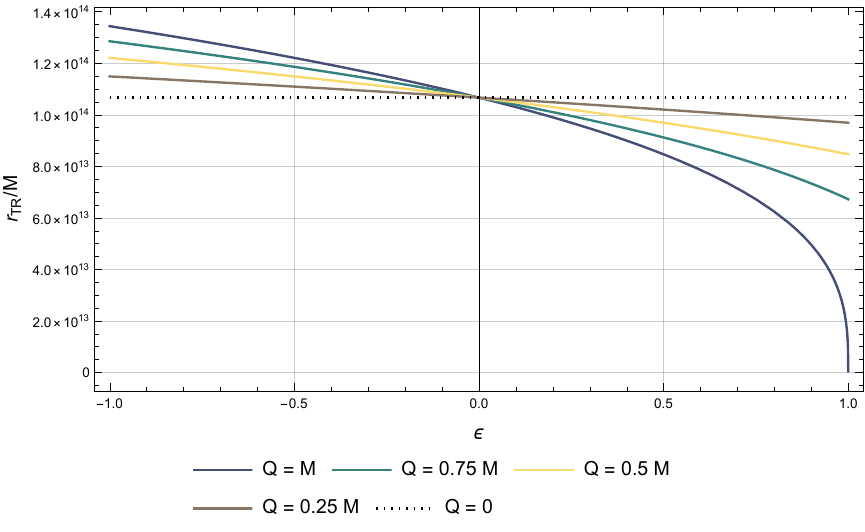}
        \caption{Turnaround radius against charge per unit mass of a charged particle that interacts with the electric field in the RNdS spacetime, with $M=1.5\times 10^5 M_\odot$, $E=1$, $\Lambda=1.1\times10^{-52} \mathrm{m}^{-2}$. For negative values of $Q$ these curves are symmetrically flipped along the y-axis.}
    \label{fig:ChargeDependance}
\end{figure}

\begin{figure}
    \centering
    \includegraphics{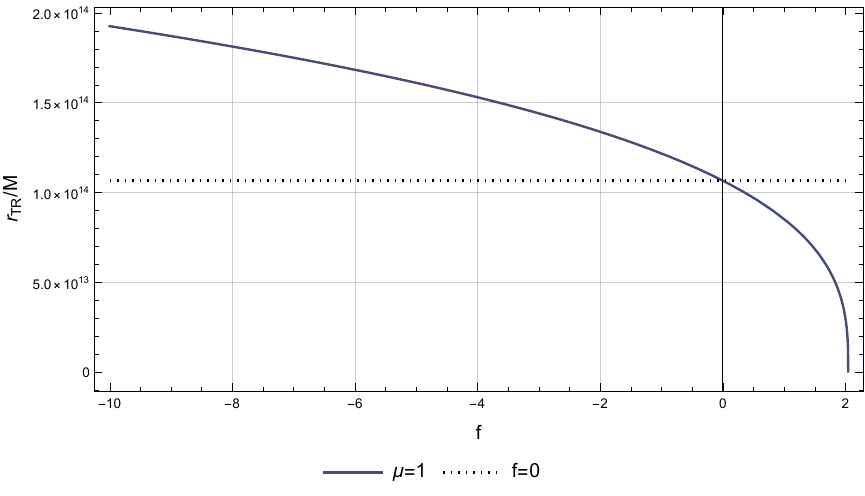}
    \caption{Turnaround radius against scalar charge coupling stregth $f$ for a test  particle that interacts with the scalar field $\varphi$, with $M=1.5\times10^5 M_\odot$, $E=1$, $\Lambda=1.1\times10^{-52} \mathrm{m}^{-2}$.}
    \label{fig:ScalarChargeDependance}
\end{figure}

Dependence on the interaction strength is illustrated in Fig.~\ref{fig:ChargeDependance} and Fig.~\ref{fig:ScalarChargeDependance} for electric and scalar interaction respectively. In the electric case the electric field contribution is quite clear. Intuition from classical electrodynamics tells us that (un)like charges (attract)repel, this translates to a turnaround radius that is (greater)less  than the turnaround radius in the uncharged case. This feature is seen rather clearly in Fig.~\ref{fig:ChargeDependance}. As the electric charge of the particle increases the electrostatic force due to the particle's interaction with the RNdS black hole increases, resulting in a decrease in the turnaround radius, until the electrostatic interaction overcomes the gravitational pull of the RNdS black hole. This happens when $\epsilon \geq \frac{M}{EQ}$.

In the scalar case, positive/negative $f$ decreases/increases the attraction of the particle towards the black hole thereby reducing/increasing the turnaround radius respectively. This is seen in Fig.~\ref{fig:ScalarChargeDependance}. As $f$ increases the turnaround decreases, until it reaches $f=2\frac{M\mu}{E^2}\sqrt{\frac{\pi}{3}}$ where the the turnaround radius vanishes. This happens as the repulsive force imparted by the scalar field interaction overcomes the gravitational attraction created by the MTZ black hole. 

\begin{figure}
    \centering
    \includegraphics{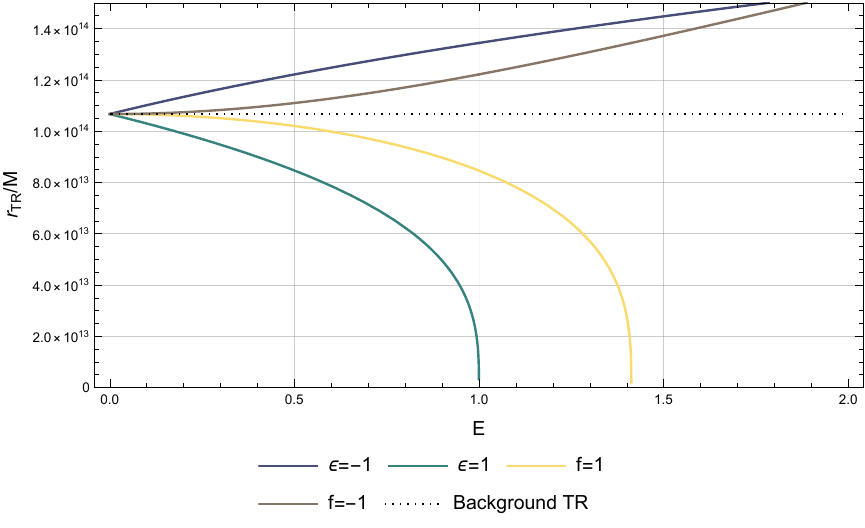}
    \caption{Turnaround radius against particle energy for particles interacting with the electric and scalar field. Here $M = 1.5\times10^5 M_\odot$, $Q= M$, and $\mu = 6.5\times10^{-6}$. This value of $\mu$ was chosen so that the the behaviour of the $f=1$ can be displayed on the same plot as the rest.} 
    \label{fig:EnergyDependance}
\end{figure}

Dependence on energy for both the electric and scalar case is illustrated in Fig.~\ref{fig:EnergyDependance}. Increasing the energy in the system increases the strength of the interaction and hence changes the turnaround radius. In the electric case the turnaround radius vanishes as $E\rightarrow \frac{M}{\epsilon Q}$, and similarly in the scalar case the turnaround radius vanishes as $E^2 \rightarrow \frac{2 M \mu}{f} \sqrt{\frac{\pi}{3}}$.
\subsection{Comparing scalar and electrically charged trajectories}
To test numerically the validity of the results, simulations of trajectories were run. Consider a particle starting at rest at position $r_0$. The motion of particles are governed by the effective potential equation $\frac{1}{2}\dot{r}^2 + V(r) =\frac{E^2}{2}$. The square on the radial velocity term creates issues in numerical analysis, to circumvent this we take a derivative with respect to proper time giving the differential equation $\ddot{r} = - \dv{V}{r}$. We then solve this with the initial conditions using Mathematica's \verb|NDSolve[]|. We use initial conditions $r(0)=r_0$ and $\dot{r}(0) = 0$, where $r_0$ is the position of the particle, chosen to be slightly below, and slightly above the corresponding turnaround radius. We also take $E=1$.

In the scalar charge case it is not so straightforward as the effective potential is derived with respect to affine parameter $\lambda$ as opposed to proper time. This gives us $r''=-\dv{V}{r}$ as our equation of motion. Integrating this differential equation will give us trajectories with respect to $\lambda$ as opposed to proper time $\tau$. To be able to compare with the electric case we must recast this into a differential equation in proper time. Using the chain rule we get that:
\begin{equation}
    r'' = \ddot{r} \left(\dv{\tau}{\lambda}\right)^2 + \dot{r} \dv[2]{\tau}{\lambda} 
\end{equation}
One can show that: 
\begin{equation}
    \dv[2]{\tau}{\lambda} = \frac{f}{\mu} \dot{r} \left(\frac{\mu + f \varphi(r)}{\mu}\right) \dv{\varphi}{r}
\end{equation}
So the equation of motion for the scalar charge particle in proper time is given to be:
\begin{equation}
    \ddot{r} \frac{(\mu + f \varphi)^2}{\mu^2} + \dot{r}^2 \frac{f}{\mu}  \left(\frac{\mu + f \varphi(r)}{\mu}\right) \dv{\varphi}{r} = - \dv{V}{r}
\end{equation}
As in the electric case we use initial conditions  $r(0) = r_0$ and $\dot{r}(0) =0$ and we only consider trajectories with $E=1$.
\begin{figure}
    \centering
    \includegraphics{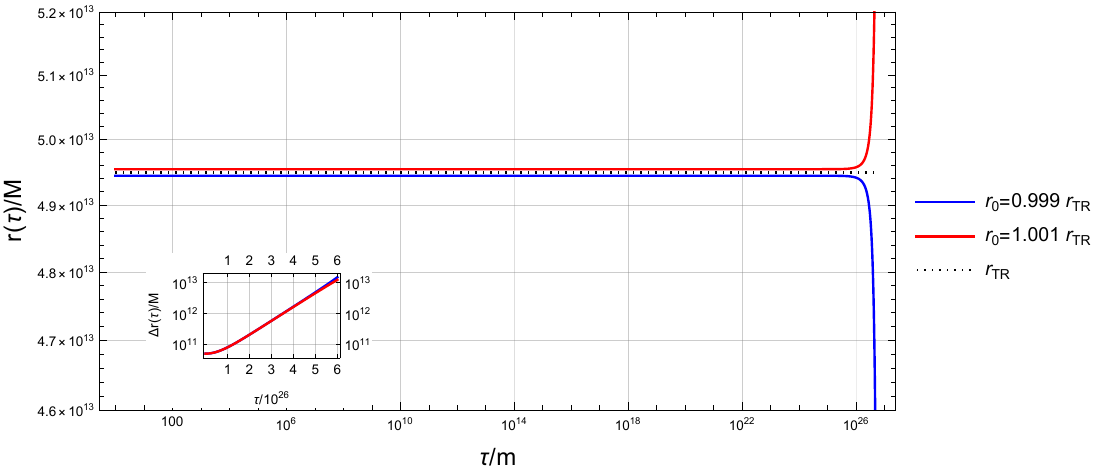}
    \caption{Trajectory of a charged particle with $\epsilon = 0.9$ in a RNdS spacetime with $Q=M=1.5\times 10^5 M_\odot$. The insert shows $\Delta r (\tau) = |r_{TR} - r(\tau)|$ on a log scale.}
    \label{fig:chargedTrajectory}
\end{figure}
\begin{figure}
    \centering
    \includegraphics{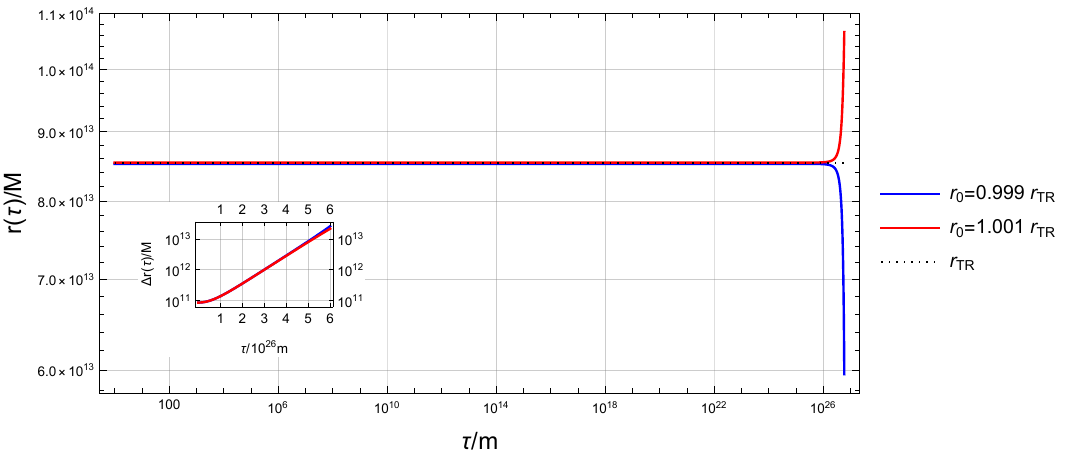}
    \caption{Trajectory of a scalar test charge with $f = 1$ and $\mu =1$ in the MTZ spacetime with $M=1.5\times10^5 M_\odot$. The turnaround radius here corrisponds to the root $r^\text{TR}_4$ in Eq.~\eqref{eqn:TRScalarChargeRoot4}.}
    \label{fig:scalarChargedTrajectory}
\end{figure}
Fig.~\ref{fig:chargedTrajectory} shows the trajectory of a charged particle in the RNdS spacetime with $Q=M=1$ and $\epsilon=0.9$. Fig.~\ref{fig:scalarChargedTrajectory} shows the trajectory of a scalar test charge with $f = 1$ in the MTZ spacetime with $M=1$. As expected, in both cases the particle just above the turnaround radius is pushed away from the black hole with the cosmic expansion, whereas the particle just below the turnaround radius eventually falls into the central black hole. One thing to note is that very close to the turnaround radius the particle stays almost stationary for a very long period of time. Should the particle start at the turnaround radius, it will remain there for all time. This is because if $r_0=r_{TR}$ then $\dd V/\dd r = 0$, which means $\ddot{r}(0) = 0$, hence the particle experiences no acceleration.

In Fig.~\ref{fig:chargedTrajectory} and Fig.~\ref{fig:scalarChargedTrajectory} we can see the deviation from the turnaround radius in both trajectories is exponential, as the log plot shows a straight line. Fitting a straight line to these log plots gives a good estimation of the lyapunov exponent of the system. For such a system the lypunov exponent can be approximated through the second derivative of the effective potential as shown in appendix \ref{sec:Lyapunov}. Evaluating this for the setup in Fig~\ref{fig:chargedTrajectory} gives a value of $k\approx 1.0488\times 10^{-26} \text{m}^{-1}$. Fitting a line to the ingoing, and outgoing trajectories in figure \ref{fig:chargedTrajectory} gives a value of $k\approx1.05209\times10^{-26} \text{m}^{-1}$ and $k\approx 1.02503\times10^{-26} \text{m}^{-1}$ respectively. These have $0.31\%$ and $2.26\%$ error respectively when compared to the analytical approximation. The numbers obtained in the scalar case when using the setup in figure \ref{fig:scalarChargedTrajectory} are identical to the charged case. This is not a coincidence. Analytically the lyapunov exponent is given as $k = \sqrt{-V''(r_\text{TR})}$ in the electric case, and $k = \sqrt{-V''(r_\text{TR})/\mathfrak{A}}$ in the scalar case, where $\mathfrak{A} = \left(\dv{\tau}{\lambda}\right)^2$ as in Eq.~\eqref{eq:dtaudalabdasquared}. If we Taylor expand these expressions about the turnaround radius, to first order in $\Lambda$, both of these expressions reduce down to $k = \sqrt{\Lambda}$. 

\section{Concluding remarks}
In this paper we have obtained the position of the turnaround radius for test electric charges in the RNdS spacetime and for test scalar charges in the MTZ spacetime. These test charges interact with the external electric field or scalar field of the spacetime and therefore move along radial non-geodesic trajectories. Therefore as expected their trajectories and the position of the turnaround radius in these spacetimes are different than those of neutral particles.  Of particular relevance is the fact that unlike the neutral case, the position of $r_{TR}$ now depends also on the parameters of the test charges themselves (such as the constant of motion $E$) besides the other parameters of the underlying spacetime. This would make sense considering that the turnaround radius is itself dependent on the non-gravitational interaction between these test charges and the background spacetimes. So through various numerical examples we have analyzed the dependence of $r_{TR}$ on these parameters. As expected the trajectories and position of $r_{TR}$ would reduce to the neutral case when the non-gravitational interaction is set to zero.

\section*{Acknowledgements}
E.J.G. would like to thank Sam Dolan for useful comments and advice. E.J.G. acknowledges financial support from  STFC.

\appendix
\section{Perturbation Expansion of Roots}
\label{appendix:PertExp}
The values of the turnaround radius are given in terms of roots of polynomials. In the uncharged, and electrically charged case these polynomials are of degree 4, whereas in the scalar charged case the polynomial is of degree 7. Although expressions exists for the exact roots of degree 4 polynomials, these are rather long; whereas in the degree 7 case such equations are non-existent. Thus we seek approximations to the values of the roots of these equations. 

The value of the cosmological constant we are using is very small, on the order of $10^{-52}$ m$^{-2}$. One may notice that in the limit $\Lambda\rightarrow0$ the polynomials simplify making the roots of the resulting polynomial much easier to calculate. This is a classical example of a perturbation theory problem. Specifically, all three polynomials fall into the singular perturbation problem category. This is because the term of highest degree in the polynomial disappears in the $\Lambda \rightarrow 0$ limit, which means that in this limit roots are `generated' by the perturbation. 

To deal with this we us the \emph{method of dominant balance} \cite{hinch1991perturbation}. This involves balancing the term of highest degree, $\Lambda r^4$ in the uncharged and electric charged case and $\Lambda r^7$ in the scalar charged case, with the term that dominates (grows the fastest) in the limit as $\Lambda\rightarrow0$. In the uncharged and electric charged cases the term proportional to $r$ is dominant, where as in the  scalar charged case the term proportional to $r^4$ dominates. Balancing gives a similarity relation which allows the definition of a change of variable, mapping our singular perturbation problem to a regular perturbation problem. 

Here in the Uncharged and electric charged cases we calculate up to a second order perturbation, and calculate up to first order in the scalar case.
\subsection{Uncharged case}
In the uncharged case we want to find the roots of the equation 
\begin{equation}
    \Lambda r^4 - 3 M r + 3 Q^2=0 \label{eqn:Uncharged polynomial}
\end{equation}
This is a singular perturbation as in the limit $\Lambda\rightarrow0$ three of the polynomial's roots escape to infinity leaving us only with one root. Applying the method of dominant balance we balance $\Lambda r^4$ with the dominant term $-3Mr$ giving us $r\sim \Lambda^{-1/3}$. Given this similarity relation we define the transformation $\xi = r \Lambda^{1/3}$. Substituting in to Eq.~\ref{eqn:Uncharged polynomial} we get:
\begin{equation}
    \xi ^4-3 M \xi +3 Q^2 \gamma =0 \label{eqn:rescaledUnchargedPol}
\end{equation}
where $\gamma = \Lambda^{1/3}$. We now expand perturbatively in $\gamma$ using the expansion:
\begin{equation}
    \xi (\gamma) = \xi_0 + \xi_1 \gamma + \xi_2 \gamma^2 + \mathcal{O}(\gamma^3) \label{eqn:UnchargedPerturbation}
\end{equation}
Substituting Eq.~\ref{eqn:UnchargedPerturbation} into Eq.~\ref{eqn:rescaledUnchargedPol}, and only keeping terms less than $\mathcal{O}(\gamma^3)$ we get:
\begin{equation}
   -3M\xi_0+\xi_0^4 + \gamma (3 Q^2 - 3 M \xi_1 + 4 \xi_0^3 \xi_1) + \gamma^2 (6 \xi_0^2 \xi_1^2 - 3 M \xi_2 + 4 \xi_0^3 \xi_2)=0
\end{equation}
We compare coefficients of $\gamma$ to obtain expressions for $\xi_i$:
\begin{align}
    \xi_0^4 -3M \xi_0 &=0\\
    3 Q^2 - 3 M \xi_1 + 4 \xi_0^3 \xi_1 &= 0\\
6 \xi_0^2 \xi_1^2 - 3 M \xi_2 + 4 \xi_0^3 \xi_2 &=0
\end{align}
Solving these for $\xi_i$'s gives 4 solutions for $\xi(\gamma)$:
\begin{equation}
    \xi(\gamma) = \begin{cases}
        \frac{Q^2 }{M} \gamma + \mathcal{O}(\gamma^3) \\
        \frac{1}{9} \gamma ^2 \left(\frac{Q^4}{\sqrt[3]{3} M^{7/3}}-\frac{i \sqrt[6]{3} Q^4}{M^{7/3}}\right)-\frac{\gamma  Q^2}{3 M}-\sqrt[3]{-3} \sqrt[3]{M}+\mathcal{O}(\gamma^3)\\
        -\frac{2 \gamma ^2 Q^4}{9 \sqrt[3]{3} M^{7/3}}-\frac{\gamma  Q^2}{3 M}+\sqrt[3]{3} \sqrt[3]{M}+\mathcal{O}(\gamma^3)\\
        \frac{2 \sqrt[3]{-\frac{1}{3}} \gamma ^2 Q^4}{9 M^{7/3}}-\frac{\gamma  Q^2}{3 M}+(-1)^{2/3} \sqrt[3]{3} \sqrt[3]{M}+\mathcal{O}(\gamma^3)
    \end{cases}
\end{equation}
transforming back to the values of $r$ and substituting $\gamma = \Lambda^{1/3}$ we get approximations for the four roots of Eq.~\ref{eqn:Uncharged polynomial}:
\begin{equation}
    r(\Lambda) = \begin{cases}
        \frac{Q^2}{M} + \mathcal{O}(\Lambda^{2/3})\\
        -\frac{Q^2}{3 M} -\frac{\sqrt[3]{-3} \sqrt[3]{M}}{\sqrt[3]{\Lambda }} + \frac{\left(\sqrt{3}-3 i\right) \sqrt[3]{\Lambda } Q^4}{9\ 3^{5/6} M^{7/3}}  +\mathcal{O}(\Lambda^{2/3})\\
        -\frac{Q^2}{3 M} +\frac{(-1)^{2/3} \sqrt[3]{3} \sqrt[3]{M}}{\sqrt[3]{\Lambda }} + \frac{2 \sqrt[3]{-\frac{1}{3}} \sqrt[3]{\Lambda } Q^4}{9 M^{7/3}}+ \mathcal{O}(\Lambda^{2/3})\\
-\frac{Q^2}{3 M}+\frac{\sqrt[3]{3} \sqrt[3]{M}}{\sqrt[3]{\Lambda }} -\frac{2 \sqrt[3]{\Lambda } Q^4}{9 \sqrt[3]{3} M^{7/3}}+ \mathcal{O}(\Lambda^{2/3})
    \end{cases}
\end{equation}
\subsection{Electric Charged case}

In the electrically charged case we find the roots to the following polynomial:
\begin{equation}
     \Lambda  r^4 +3 rE Q \epsilon -3 Mr -3 Q^2 \left(\epsilon ^2-1\right) =0.
\end{equation}
This polynomial still has 4 roots. It also has the same term structure as Eq.~\ref{eqn:Uncharged polynomial}, in that it is a singular perturbation problem with polynomial of degree 4 and only has terms proportional to $r$ and a constant term. Because of this procedure of applying the method of dominant balance will be identical. We can immediately transform to $\xi = r \Lambda^{1/3}$. This gives the equivalent formulation
\begin{equation}
         \xi^4 +(3  E Q \epsilon -3 M)\xi -3\gamma Q^2 \left(\epsilon ^2-1\right) = 0\label{eqn:rescaledChargedPol}
\end{equation}
where again $\gamma=\Lambda^{1/3}$. We expand to second order in $\gamma$:
\begin{equation}
    \xi (\gamma) = \xi_0 + \xi_1 \gamma + \xi_2 \gamma^2 + \mathcal{O}(\gamma^3) \label{eqn:ChargedPerturbation}
\end{equation}
Substituting Eq.~\ref{eqn:ChargedPerturbation} into Eq.~\ref{eqn:rescaledChargedPol}, and only keeping terms of order less than $\mathcal{O}(\gamma^2)$ we get:
\begin{align}
    \nonumber -3 M \xi_0+\gamma  \left(-3 M \xi_1+4 \xi_0^3 \xi_1-3 Q^2 \epsilon ^2+3 Q^2+3 E\xi_1 Q \epsilon \right)\\+\xi_0^4+3 E \xi_0 Q \epsilon
    +\gamma^2 (6 \xi_0^2 \xi_1^2 - 3 M \xi_2 + 3 Q \epsilon E \xi_2 + 4 \xi_0^3 \xi_2) = 0
\end{align}
Solving for the $\xi_i$'s and then transforming back to obtain solutions for $r$ gives:
\begin{equation}
    r(\Lambda)=\begin{cases}
        \frac{Q^2 (\epsilon^2 -1)}{E Q \epsilon -M} + \mathcal{O}(\Lambda^{2/3}) \\
\frac{Q^2 (\epsilon^2 -1)}{3 (M-E Q \epsilon )}-\frac{\sqrt[3]{-3} \sqrt[3]{M-E Q \epsilon }}{\sqrt[3]{\Lambda }}-\frac{2 (-1)^{2/3} \sqrt[3]{\Lambda } Q^4 (\epsilon^2 -1)^2}{9 \sqrt[3]{3} (M-E Q \epsilon )^{7/3}} + \mathcal{O}(\Lambda^{2/3})\\

\frac{Q^2 (\epsilon^2 -1)}{3 (M-E Q \epsilon )}+\frac{(-1)^{2/3} \sqrt[3]{3} \sqrt[3]{M-E Q \epsilon }}{\sqrt[3]{\Lambda }} + \frac{2 \sqrt[3]{-\frac{1}{3}} \sqrt[3]{\Lambda } Q^4 (\epsilon^2 -1)^2}{9 (M-E Q \epsilon )^{7/3}} + \mathcal{O}(\Lambda^{2/3})\\

\frac{Q^2 (\epsilon^2 -1)}{3 (M-E Q \epsilon )}+\frac{\sqrt[3]{3} \sqrt[3]{M-E Q \epsilon }}{\sqrt[3]{\Lambda }}-\frac{2 \sqrt[3]{\Lambda } Q^4 (\epsilon^2 -1)^2}{9 \sqrt[3]{3} (M-E Q \epsilon )^{7/3}} + \mathcal{O}(\Lambda^{2/3})

    \end{cases} 
\end{equation}
\subsection{Scalar Charged case}
Written out as a polynomial only in $r$ and $\Lambda$, Eq.~\eqref{eqn:TRScalarCharge} can bee written as:
\begin{equation}
    a_7 \Lambda r^7 + a_6 \Lambda r^6 + a_5 \Lambda r^5 + a_{4\Lambda} \Lambda r^4 + a_4 r^4 + a_3r^3 + a_2 r^2 + a_1r + a_0 = 0, 
\end{equation}
where the $a_i$'s are the other coefficients of the polynomial, depending on $f,\mu, M$ and $E$. This is also a singular perturbation problem as the $r^7$ vanishes in the $\Lambda\rightarrow0$ limit. The only way to balance the dominant term with $\Lambda r^7$ is by setting $\Lambda r^7 \sim r^4$ giving $r\sim \Lambda^{-1/3}$. With this we can define $\xi = r \Lambda^{1/3}$. Applying this transformation to the polynomial we get a polynomial of the form: 
\begin{equation}
    a_7 \xi^7 + a_6 \gamma \xi^6 + a_5 \gamma^2 \xi^5 +  a_{4\Lambda}\gamma^3 \xi^4  + a_4 \xi^4 + a_3 \gamma \xi^3 +  a_2 \gamma^2 \xi^2 + a_1 \gamma^3 \xi + \gamma^4 a_0 =0,
\end{equation}
where $\gamma = \Lambda^{1/3}$. We perturb linearly in $\gamma$:
\begin{equation}
    \xi(\gamma) = \xi_0 + \xi_1 \gamma  + \mathcal{O}(\gamma^2).
\end{equation}
Substituting this perturbation into the polynomial gives a long expression with terms up to $\mathcal{O}(\gamma^ 7)$. Setting each coefficient of $\gamma$ to 0 gives the following 7 roots:
\begin{equation}
    r(\Lambda) = \begin{cases}
    \frac{M \left(-9 E^2 f^2+4 \sqrt{3 \pi } E^2 f \mu +4 \pi  \mu ^2\right)}{6 \mu  \left(\sqrt{3 \pi } E^2 f-2 \pi  \mu \right)}-\frac{\sqrt[3]{-\frac{3}{2 \pi }} \sqrt[3]{M \left(2 \pi  \mu -\sqrt{3 \pi } E^2 f\right)}}{\sqrt[3]{\Lambda } \sqrt[3]{\mu }} + \mathcal{O}(\Lambda^{1/3})\\
    \frac{M \left(-9 E^2 f^2+4 \sqrt{3 \pi } E^2 f \mu +4 \pi  \mu ^2\right)}{6 \mu  \left(\sqrt{3 \pi } E^2 f-2 \pi  \mu \right)}+\frac{(-1)^{2/3} \sqrt[3]{3 \mu  M-\frac{3}{2} \sqrt{\frac{3}{\pi }} E^2 f M}}{\sqrt[3]{\Lambda } \sqrt[3]{\mu }} + \mathcal{O}(\Lambda^{1/3})\\
    \frac{M \left(-9 E^2 f^2+4 \sqrt{3 \pi } E^2 f \mu +4 \pi  \mu ^2\right)}{6 \mu  \left(\sqrt{3 \pi } E^2 f-2 \pi  \mu \right)}+\frac{\sqrt[3]{3 \mu  M-\frac{3}{2} \sqrt{\frac{3}{\pi }} E^2 f M}}{\sqrt[3]{\Lambda } \sqrt[3]{\mu }} + \mathcal{O}(\Lambda^{1/3})\\
    M + \mathcal{O}(\Lambda^{1/3})\\
    r_5 + \mathcal{O}(\Lambda^{1/3})\\
    r_6 + \mathcal{O}(\Lambda^{1/3})\\
    r_7 + \mathcal{O}(\Lambda^{1/3})
    \end{cases}
\end{equation}
with $r_5, r_6, r_7$ being very long expressions given in Appendix \ref{appendix:abc}.
\section{The Roots $r_5, r_6, r_7$ in the scalar case.}
\label{appendix:abc}
\begin{figure}
        \centering
     \begin{subfigure}[b]{0.45\textwidth}
         \centering
         \includegraphics[width=\textwidth]{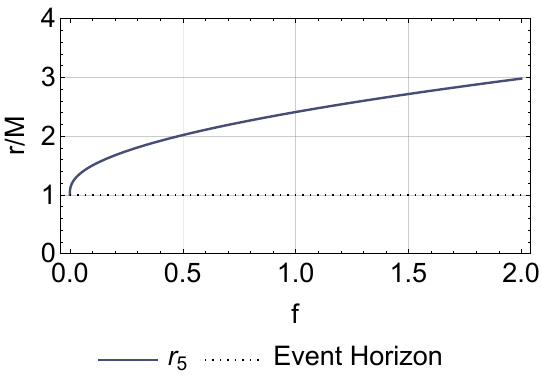}
         \caption{Plot of $r_5$ against coupling strength $f$. }
          \label{fig:r5SC}
     \end{subfigure}
     \hfill
     \begin{subfigure}[b]{0.45\textwidth}
         \centering
         \includegraphics[width=\textwidth]{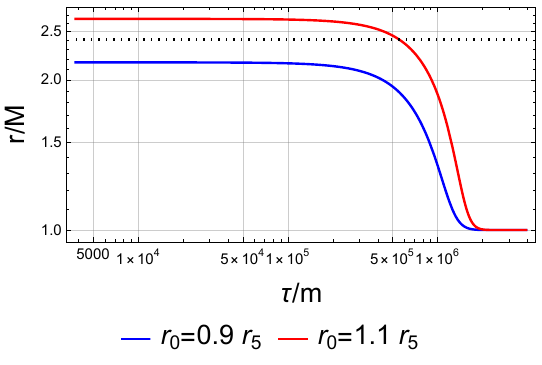}
         \caption{Plot of Trajectories starting above and below $r_5$ for $f=1$. The dotted line indicates the location of $r_5$.}
         \label{fig:r5Trajectories}
     \end{subfigure}
        \caption{Plots analysing the behaviour of the root $r_5$. Here we take $M=10^5 M_\odot$, $E=1$ and $\mu =1$.}
        \label{fig:r5Graphs}
    \end{figure}

        \begin{align}
        \mathcal{A}^3 =& -9 \sqrt{3 \pi } E^4 f^5 \mu ^4 M^3+18 \pi  E^2 \left(3 E^2-1\right) f^4 \mu ^5 M^3\\
        &-36 \sqrt{3} \pi ^{3/2} E^2 \left(E^2-1\right) f^3 \mu ^6 M^3+24 \pi ^2 E^2 \left(E^2-3\right) f^2 \mu ^7 M^3\\
        &+\sqrt{3 \pi } \mathcal{B} +16 \pi ^{5/2} \sqrt{3} E^2 f \mu ^8 M^3\\
        \mathcal{B}^2 =& \  E^4 f^2 \mu ^8 M^6 \left(81 E^4 f^8-108 \sqrt{3 \pi } E^2 \left(3 E^2+1\right) f^7 \mu  \right. \\
        &+108 \pi  \left(15 E^4+12 E^2+1\right) f^6 \mu ^2-144 \sqrt{3} \pi ^{3/2} \left(10 E^4+15 E^2+3\right) f^5 \mu ^3\\
        &+720 \pi ^2 \left(3 E^4+8 E^2+3\right) f^4 \mu ^4-192 \sqrt{3} \pi ^{5/2} \left(3 E^4+15 E^2+10\right) f^3 \mu ^5\\
        & \left. +192 \pi ^3 \left(E^4+12 E^2+15\right) f^2 \mu ^6-256 \sqrt{3} \pi ^{7/2} \left(E^2+3\right) f \mu ^7+256 \pi ^4 \mu ^8\right) 
        \end{align}
        \begin{align}
        r_5 =& -\frac{\mathcal{A}}{2 \mu ^2 \left(\sqrt{3 \pi } E^2 f-2 \pi  \mu \right)}
        -\frac{E^2 f \mu  M^2 \left(3 \sqrt{3} f^2-12 \sqrt{\pi } f \mu +4 \pi  \sqrt{3} \mu ^2\right)}{\left(\sqrt{3} E^2 f-2 \sqrt{\pi } \mu \right) \mathcal{A}}
        +\frac{M \left(\sqrt{3 \pi } f-2 \pi  \mu \right)}{\sqrt{3 \pi } E^2 f-2 \pi  \mu }
        \end{align}
        \begin{align}
        r_6 = \frac{\left(1-i \sqrt{3}\right) \mathcal{A}}{4 \mu ^2 \left(\sqrt{3 \pi } E^2 f-2 \pi  \mu \right)}+\frac{\left(1+i \sqrt{3}\right) E^2 f \mu  M^2 \left(3 \sqrt{3} f^2-12 \sqrt{\pi } f \mu +4 \pi  \sqrt{3} \mu ^2\right)}{2 \left(\sqrt{3} E^2 f-2 \sqrt{\pi } \mu \right) \mathcal{A}}+\frac{M \left(\sqrt{3 \pi } f-2 \pi  \mu \right)}{\sqrt{3 \pi } E^2 f-2 \pi  \mu }
        \end{align}
        \begin{align}
        r_7 = \frac{\left(1+i \sqrt{3}\right) \mathcal{A}}{4 \mu ^2 \left(\sqrt{3 \pi } E^2 f-2 \pi  \mu \right)}+\frac{\left(1-i \sqrt{3}\right) E^2 f \mu  M^2 \left(3 \sqrt{3} f^2-12 \sqrt{\pi } f \mu +4 \pi  \sqrt{3} \mu ^2\right)}{2 \left(\sqrt{3} E^2 f-2 \sqrt{\pi } \mu \right) \mathcal{A}}+\frac{M \left(\sqrt{3 \pi } f-2 \pi  \mu \right)}{\sqrt{3 \pi } E^2 f-2 \pi  \mu }
        \end{align}
    These equations are rather complicated, so here we only do some simple analysis. $r_6$ and $r_7$ are clearly a complex conjugate pair and therefore nonphysical. $r_5$ however is possibly a physical root. Investigating this root numerically for some given parameters gives Fig.~\ref{fig:r5SC}. One may note for the chosen parameters $r_5$ is indeed physical, i.e. real and outside the event horizon. $r_1$ and $r_4$ are minima and maxima in the effective potential respectfully, so the only possible stationary point between these two is a point of inflection in the effective potential. Running some simulations for particles starting above and below this point, as in  Fig.~ \ref{fig:r5Trajectories}, shows that particles starting above and below this point will both fall below the event horizon. This behaviour is characteristic of an inflection point.

\section{Analytical Approximation of the Lyapunov exponent.}
\label{sec:Lyapunov}
In this section approximate analytical formulae for the lyapunov exponent for trajectories moving away from the turnaround radius are obtained, in both the electric case, and the scalar case. 
\subsection{Electric case.}
In the electric case the equation of motion is $\ddot{r} = -\dv{V}{r}$. We Taylor expand the effective potential function about the turnaround radius, given that at this radius the first derivative of the effective potential is zero:
\begin{equation}
    V(r) \approx V(r_{TR}) + \frac{1}{2} V''(r_{TR}) (r-r_TR)^2
\end{equation}
We can then approximate the derivative near the turnaround radius by:
\begin{equation}
    V'(r) \approx V''(r_{TR}) (r-r_{TR}) \label{eqn:approxEffPot}
\end{equation}
To find the lyapunov exponent we assume the behaviour of the function is simple harmonic and thus $\ddot{r} = k^2 (r-r_{TR})$ and $\dot{r} = k(r-r_{TR})$, substituting this, and using Eq.~\ref{eqn:approxEffPot} the equation of motion becomes: 
\begin{equation}
    k^2 (r-r_{TR}) + V''(r_{TR}) (r-r_{TR}) = 0
\end{equation}
Thus we can conclude that the maximal lyapunov exponent will be given by 
\begin{equation}
    k = \sqrt{-V''(r_{TR})}
\end{equation}
\subsection{The Scalar Case.}
In the scalar case the equation of motion is of the form:
\begin{equation}
    \mathfrak{A}(r) \ddot{r} + \mathfrak{B}(r) \dot{r}^2 + \dv{V}{r} = 0
\end{equation}
where $\mathfrak{A}$ and $\mathfrak{B}$  are functions of r. 
Again we Taylor expand about the turnaround radius in the same manor as Eq.~\eqref{eqn:approxEffPot}. Substituting that and $\ddot{r} = k^2 (r-r_{TR})$ and $\dot{r}=k(r-r_{TR})$, only keeping terms of zeroth order in $(r-r_{TR})$ we get that:
\begin{equation}
    \mathfrak{A}(r_{TR}) k^2 + V''(r_{TR}) = 0
\end{equation}
which gives the maximal lyapunov exponent:
\begin{equation}
    k = \sqrt{-\frac{V''(r_{TR})}{\mathfrak{A}(r_{TR})}}
\end{equation}
\bibliographystyle{apsrev4-1}
\bibliography{ref.bib}
\end{document}